\documentclass[a4paper,twocolumn,showpacs,amsmath,pra,amssymb,nofootinbib]{revtex4-1}  
\usepackage[T1]{fontenc}
\usepackage[latin9]{inputenc} 
\usepackage{times}
\usepackage{color}
\usepackage{xspace}
\usepackage{amssymb,amsmath}
\usepackage{amsbsy}
\usepackage{graphicx}  %

\newcommand{\UD}{U_{\mathrm{dd}}}
\newcommand{\UDt}{\tilde{U}_{\mathrm{dd}}}
\newcommand{\Cdd}{C_{\mathrm{dd}}}
\newcommand{\br}{\mathbf{r}}
\newcommand{\bx}{\mathbf{x}}
\newcommand{\bk}{\mathbf{k}}
\newcommand{\ldb}{\lambda_{\mathrm{dB}}}
\newcommand{\kt}{k_BT}
\newcommand{\od}[1]{\ensuremath{{O\negthickspace\left({{#1}}\right)}}\xspace}
\newcommand{\aho}{a_{\mathrm{ho}}\xspace}

\begin{document}
\title{Magnetostriction and exchange effects in trapped dipolar Bose and Fermi gases}
\author{D.~Baillie} 
\author{P.~B.~Blakie}  

\affiliation{Jack Dodd Centre for Quantum Technology, Department of Physics, University of Otago, Dunedin, New Zealand.}

\begin{abstract}
We examine the magnetostrictive position and momentum space distortions that occur in harmonically confined dipolar Bose and Fermi gases. Direct interactions give rise to position space magnetostriction and  exchange interactions give rise to momentum space magnetostriction. While the position space magnetostriction is similar in Bose and Fermi systems, the  momentum space magnetostriction is markedly different: the  Bose gas momentum distribution distorts in the opposite sense to that of the Fermi gas.  By relating  exchange effects to short range correlations between the particles we discuss the energetic origin of this difference.  Our main calculations are based on Hartree-Fock theory, but we also provide analytic approximations  for the magnetostriction effects at zero and finite temperature. Our predictions should be verifiable in current experiments with ultra-cold polar molecules. 
\end{abstract}


\maketitle

\section{Introduction}

Recently  a range of ultra-cold gases with appreciable magnetic \cite{Griesmaier2005a,*Bismut2010a,*Pasquiou2011a,Mingwu2011a,Aikawa2012a,Lu2012a}  and electric \cite{Aikawa2010a,Ni2008a} dipoles have been realized in experiments.  The key feature of these gases is that the constituent particles interact via a dipole-dipole interaction (DDI) that is long-ranged and anisotropic. When these systems are polarized by an external field  the DDI causes their shape to distort. The first experimental evidence for such a magnetostrictive effect\footnote{We adopt terminology appropriate to magnetic dipole gases. We note that for polar molecule gases the effects we predict are electrostrictive.} was in the expansion of   dipolar Bose-Einstein condensates (BECs) \cite{Stuhler2005a}  (also see \cite{Yi2003a,Giovanazzi2003a,Giovanazzi2006a,Mingwu2011a}).

Here we examine magnetostrictive effects in normal Bose and Fermi gases with polarized dipoles. In our work we distinguish between two cases: position space magnetostriction  and momentum space magnetostriction, whereby the position and momentum space distributions, respectively, distort relative to the direction of the polarized dipoles. We show that position space magnetostriction is qualitatively similar for Bose and Fermi gases: the gas tends to elongate along the dipole polarization direction to reduce the direct interaction energy. Momentum space magnetostriction arises solely from exchange interactions and distinguishes the quantum statistics of the particles: the Fermi gas momentum distribution elongates along the polarization direction, while the Bose gas instead contracts along this direction.
 
Previous studies have considered the distortion of ultra-cold gases with DDIs. Work on Bose gases has focused on position space magnetostriction for zero temperature BECs  \cite{Goral2000a,Goral2002a,Eberlein2005a}. For normal Fermi gases Hartree calculations (excluding exchange) were used to study position space magnetostriction in \cite{Goral2001a}. The first theories including exchange interactions for the trapped Fermi gas were based on simple variational approximations    \cite{Miyakawa2008a,Lima2010a,Lima2010b,Endo2010a} (also see \cite{Endo2011a}), and revealed distortion of the momentum distribution. 
In this paper we use full Hartree-Fock (HF) calculations to provide a unified quantitative treatment of trapped Bose and Fermi gases. While HF calculations have been applied to trapped Fermi gases \cite{Zhang2009a,Zhang2010a,Baillie2010b}, this is the first application of this theory the trapped dipolar Bose gas (c.f.~Hartree analysis reported in \cite{Bisset2011}, also see \cite{Ticknor2012a}).    Our main concern is in characterizing  magnetostrictive effects,  their relationship to  direct and exchange interactions, and to pair correlations. We support our numerical results by analytical approximations.

\section{Theory}
We consider a gas of  polarized dipolar particles that interact by a long-range dipole-dipole interaction of the form
\begin{align}
\UD(\bx)=\frac{\Cdd}{4\pi}\frac{1-3\cos^2\theta}{|\bx|^3},\label{e:udd}
\end{align}
where the constant $\Cdd$ is given by $\mu_0\mu^2_m$ for magnetic dipoles of strength $\mu_m$ and $d^2/\epsilon_0$ for electric dipoles of strength $d$, and $\theta$ is the angle between $\bx$ and the polarization axis, which we take to be the $z$ direction. The atoms are confined within a cylindrically symmetric harmonic trap  
\begin{equation}
U_{\mathrm{tr}}(\bx)=\frac{m}{2}\left[\omega_\rho^2(x^2+y^2)+\omega_z^2z^2\right],
\end{equation}
with aspect ratio $\lambda=\omega_z/\omega_\rho$. 

Following \cite{Zhang2009a,Zhang2010a,Baillie2010b,Zhang2011a} we introduce the dimensionless interaction parameter $D_t = \Cdd N^{1/6}/(4\pi\hbar\omega \aho^3)$ where $\omega=(\omega_\rho^2\omega_z)^{1/3}$ and $\aho=\sqrt{\hbar/m\omega}$.  When considering thermal effects we characterize the temperature either in terms of the ideal condensation temperature, $T_c^0=\sqrt[3]{N/\zeta(3)}\hbar\omega/k_B$, with $\zeta(s)$ the zeta function, or the Fermi temperature, $T_F^0=\sqrt[3]{6N}\hbar\omega/k_B$. We note that these two degeneracy temperatures relate as $T_c^0\approx0.52T_F^0$.

\subsection{Semiclassical theory}
In the semi-classical approximation \cite{Goral2001a,Zhang2010a,Baillie2010b} the system is described by the Wigner function  
\begin{align}
W(\bx,\bk)=\frac{1}{\exp([\epsilon(\bx,\bk)-\mu]/\kt)-\eta},\label{e:wigner}
\end{align}
where $\eta=1$ for bosons and $\eta=-1$ for fermions, and $\mu$ is the chemical potential. This approximation furnishes a good description when the temperature is large compared to the trap level spacing ($k_BT\gg \hbar\omega$). For fermions this approximation can be applied at $T=0$ as long as the Fermi energy is sufficiently large (i.e.~$\mu\gg\hbar\omega$) \cite{Zhang2009a}. 
For the case of bosons at $T<T_c$, where $T_c$ is the condensation temperature \cite{Glaum2007}, a condensate emerges and the semi-classical theory requires augmentation by a beyond-semi-classical description of the condensate. Here we restrict our attention to normal Bose and Fermi gases.

The position space and momentum space densities are given by 
\begin{align}
n(\bx)=\int  \frac{d\bk}{(2\pi)^{3}} W(\bx,\bk),\label{nx}\\  
\tilde{n}(\bk)= \int \frac{d\bx}{(2\pi)^3} W(\bx,\bk),
\end{align}
respectively and the total number is given by $N=\int d\bx \,n(\bx)$.

\subsection{Characterization of magnetostriction}
Our primary concern is distortions in the trapped clouds arising from the DDI. It is convenient to characterize this in terms of the rms-widths
\begin{align}
\sigma_{\nu}&\equiv\left[\frac{1}{N}\int \frac{d\bx d\bk}{(2\pi)^3}\nu^2W(\bx,\bk)\right]^{1/2},
\end{align}
with $\nu=\{x,y,z,k_x,k_y,k_z\}$.

In terms of these quantities we define the momentum and position distortion as 
\begin{align} 
\alpha&\equiv\frac{\sigma_{k_x}}{\sigma_{k_z}},\quad
\beta\equiv\frac{1}{\lambda}\frac{\sigma_{x}}{\sigma_{z}}, 
\end{align}
respectively, i.e.~as the ratio of the widths in the directions perpendicular ($x$ or $k_x$) and parallel ($z$ or $k_z$) to the dipolar polarization direction.
 The trap anisotropy appears in the definition of $\beta$ so that in the absence of interactions, where the position density has the  anisotropy imposed by the trap, we have $\beta=1$.  In contrast the non-interacting momentum distribution is spherically symmetric. We refer to systems as exhibiting position space magnetostriction when $\beta\ne1$ and momentum space magnetostriction when $\alpha\ne1$.

\subsection{Hartree-Fock theory\label{s:formalism}}
The HF dispersion relation is \cite{Zhang2010a,Lin2010a,Baillie2010b}
\begin{align}
\epsilon(\bx,\bk)=\frac{\hbar^2k^2}{2m}+U_{\mathrm{tr}}(\bx)+\Phi_D(\bx)+\eta\Phi_E(\bx,\bk),\label{e:epsilon}
\end{align}
with
\begin{align}
\Phi_D(\bx)&=\int \frac{d\bx' d\bk'}{(2\pi)^3}\UD(\bx-\bx')W(\bx',\bk'),\label{e:PhiD}\\
\Phi_E(\bx,\bk)&=\int \frac{ d\bk'}{(2\pi)^3}\UDt(\bk-\bk')W(\bx,\bk'),\label{e:PhiE}
\end{align}
the direct and exchange interaction terms, respectively. In Eq.~\eqref{e:PhiE}, $\UDt(\bk)=\Cdd (\cos^2\theta_\bk-1/3)$  is the Fourier transform of the dipole-dipole interaction, where $\theta_\bk$ is the angle between $\bk$ and $\hat{k}_z$.  The HF theory requires solving Eqs.~\eqref{e:wigner}, \eqref{e:epsilon}-\eqref{e:PhiE} in an iterative procedure until the equations are self-consistent. We are interested in solutions for a specified total number, $N$, which requires an additional iterative procedure to adjust $\mu$.

Our system is cylindrically symmetric, since the symmetry axis of the harmonic trap  coincides with the polarization direction. We can exploit this symmetry to simplify the Wigner function to a	function of four	variables $\{\rho,z,k_\rho,k_z\}$, i.e., cylindrical coordinates in $\bx$ and $\bk$ space. 
Our numerical algorithm makes use of Fourier transform techniques and quadrature based on discrete cosine and Bessel functions, and is detailed in \cite{Baillie2010b} (also see \cite{Zhang2010a}). The numerical implementation of HF theory requires extensive computational resources because of the necessity to construct the full Wigner function on grids with suitable resolution to accurately evaluate the DDI.

\subsection{Hartree theory} 
We note that the direct interaction potential (\ref{e:PhiD}) can be evaluated in terms of the position space density, i.e.~$\Phi_D(\bx)=\int  d\bx'\,\UD(\bx-\bx')n(\bx')$, thus if the exchange term is neglected the $\bk$ dependence of the dispersion (\ref{e:epsilon}) is  quadratic and Eq.~(\ref{nx}) can be evaluated \cite{Baillie2010b,Bisset2011}
\begin{equation}
n(\bx)=\frac{1}{\ldb^3}\zeta_{3/2}^\eta\left(e^{[\mu-V_{\mathrm{eff}}(\bx)]/k_BT}\right),\label{den}
\end{equation}
where  $\zeta_{\alpha}^{\eta}(z)=\sum_{j=1}^{\infty}\eta^{j-1}z^j/j^\alpha$ is the polylogarithm function,  $\ldb=h/\sqrt{2\pi mk_BT}$, and 
\begin{equation}
V_{\mathrm{eff}}(\bx)\equiv U_{\mathrm{tr}}(\bx)+\Phi_D(\bx),\label{Veff}
\end{equation}
is the effective potential. By neglecting exchange we have arrived at a Hartree theory for the system, which involves solving Eqs.~(\ref{den}) and (\ref{Veff}) self-consistently. This theory is computationally much simpler because it does not require the evaluation of the full Wigner function. The Hartree theory provides a baseline for investigating the exchange effects included in the full HF calculations.

\subsection{Analytic approximations}
To develop an analytic approximations for the magnetostrictive behavior we introduce the following ansatz for the HF dispersion relation:
\begin{align}
    \epsilon_A(\bx,\bk) = \frac{\hbar^2}{2m}\left( \kappa_\rho k_\rho^2  + \kappa_z k_z^2\right) +\frac{m\omega_\rho^2}{2}\left( \chi_\rho \rho^2+\chi_z \lambda^2 z^2\right),
\end{align}
where $\kappa_\rho,\kappa_z,\chi_\rho$ and $\chi_z$ are variational parameters to be determined.
\subsubsection{Zero temperature fermions}
We first consider an analytic treatment of $T=0$ fermions. For this system all single particle states up to the Fermi energy are occupied.
We use a Thomas-Fermi ansatz for the Wigner function (after \cite{Miyakawa2008a})
\begin{align}
    W_0(\bx,\bk) &= \Theta[\mu-\epsilon_A(\bx,\bk)], \label{e:W0}
\end{align}
where $\Theta[\cdot]$ is the Heaviside step function. In the appendix, we minimize the energy to determine the variational parameters, from which we obtain $\alpha$ and $\beta$. We find, to first order, that
\begin{align}
    \alpha &\approx  1 - c_\alpha D_t, \label{e:alphastep}\\
    \beta &\approx 1 + c_\beta  \lambda^{-2}J'(\lambda^{-2}-1)  D_t, \label{e:betastep}
\end{align}
where $c_\alpha =  \frac{2^{38/3}}{3^{17/6} 175 \pi^2} \approx 0.167$, $c_\beta =  \frac{2^{35/3}}{3^{11/6}35 \pi^2} \approx 1.26$, and the function $J(u)$ is defined in Eq.~(\ref{e:Ju}).
\subsubsection{Finite temperature}
To account for thermal effects in Bose and Fermi gases we use a  Boltzmann ansatz for the Wigner function  (after \cite{Sogo2009a,Endo2011a})
\begin{align}
    W_T(\bx,\bk) &= e^{\left[ \mu -\epsilon_A(\bx,\bk)\right]/\kt}.\label{e:WT} 
\end{align}
This approximation fails to account for the quantum statistics (e.g.~phase space densities greater than unity are not prohibited for Fermions), and thus is only applicable at temperatures well above the relevant degeneracy temperatures.\footnote{We are unable to develop simple closed-form analytic results using degenerate forms of the Wigner function [c.f. Eq.~\eqref{e:wigner}]. Nevertheless, such a degenerate form could be treated numerically and would lead to better agreement with the full HF results.} 
 In the appendix, we minimize the free energy\footnote{We note that Bose and Fermi statistics are included in the treatment of interactions by adding the appropriate exchange interaction term.} derived from (\ref{e:WT}) to determine the variational parameters and hence the distortion parameters. To first order we find
 \begin{align}
   \alpha &\approx 1 + \frac{\eta}{20\sqrt{\pi}} \left(\frac{\hbar\omega N^{1/3}}{\kt}\right)^{5/2} D_t   \label{e:alphagaussian},\\
   \beta &\approx 1 +\frac{3}{8\sqrt{\pi}}\lambda^{-2}  J'(\lambda^{-2}-1)  \left(\frac{\hbar\omega N^{1/3}}{\kt}\right)^{5/2} D_t.  \label{e:betagaussian}
\end{align}
\section{Results}

\subsection{Position space magnetostriction}
\begin{figure}[!tH]
\begin{center} 
\includegraphics[width=3.4in]{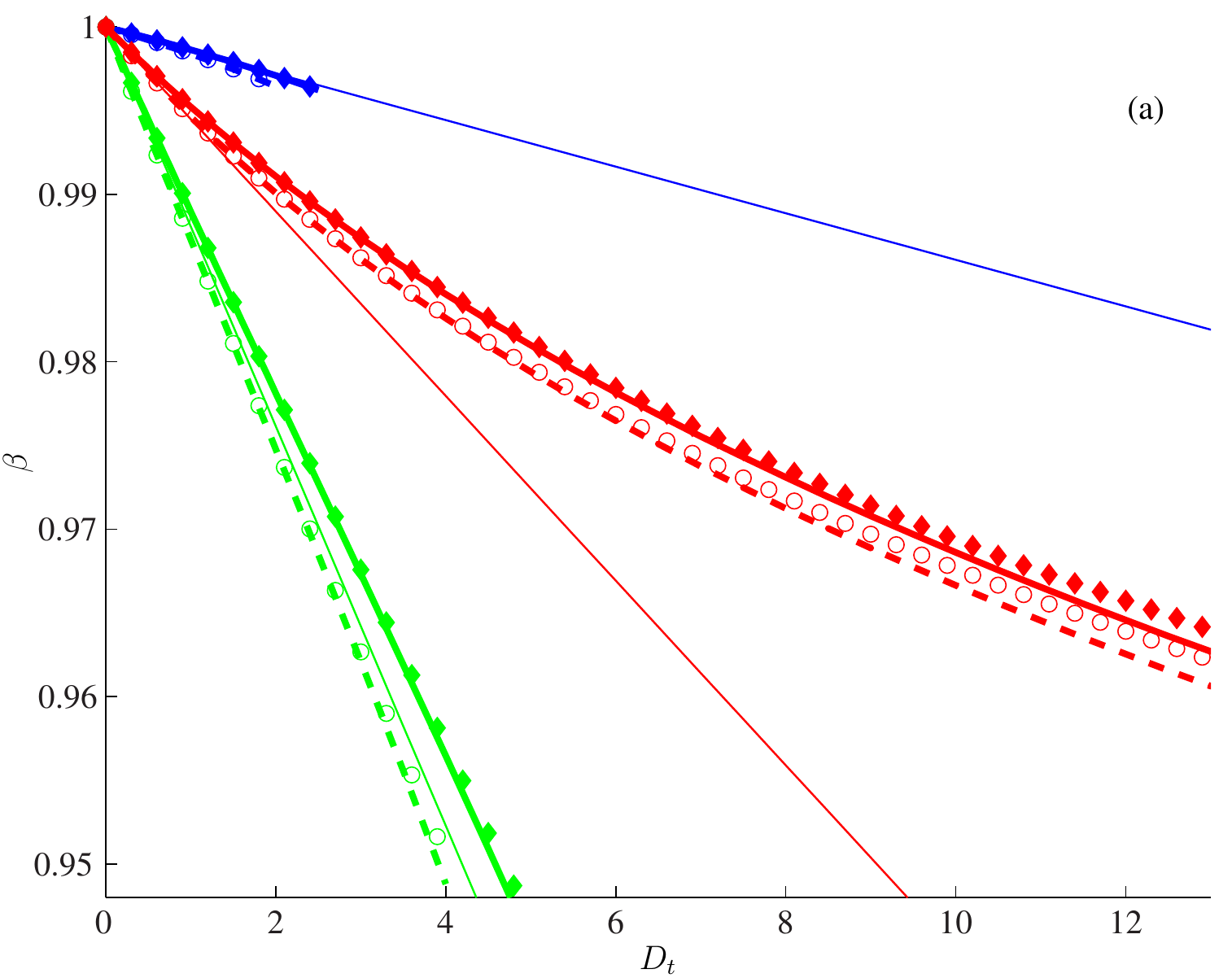} 
\includegraphics[width=3.4in]{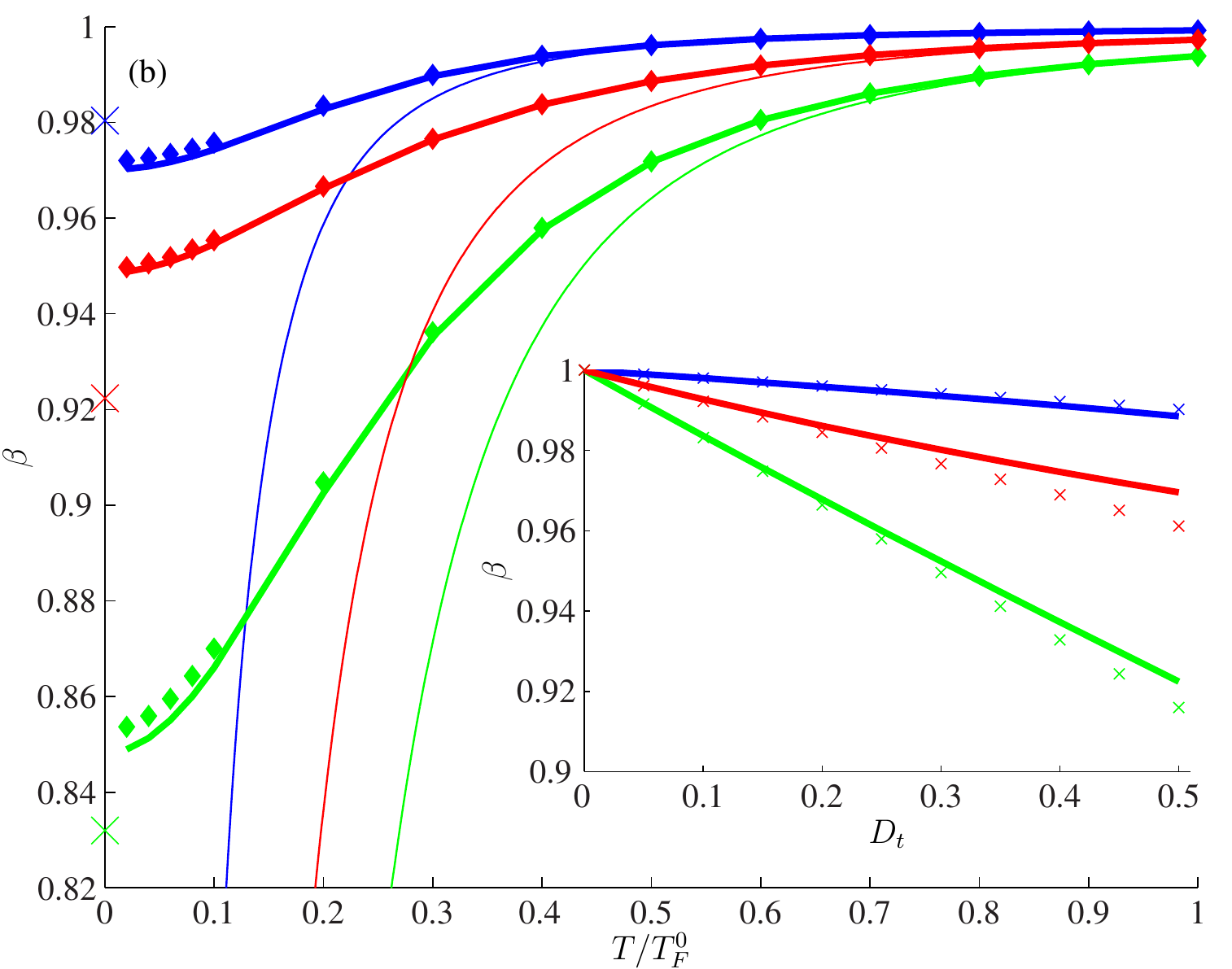}  
\caption{\label{f:beta}(Color online) Position space distortion for fermions using HF (solid curves) and Hartree (diamonds), and for bosons using HF (dashed curves) and Hartree (circles), also the predictions from   \eqref{e:betastep} (crosses) and \eqref{e:betagaussian} (thin curves) with $\lambda$ = 0.1 [blue (dark gray)], 1 [green (light gray)], and 10 [red (gray)]. (a) Bosons and fermions at $T=1.5T_c^0\approx0.78T_F^0$. (b) Fermions with $D_t=1$. (inset) Fermions at $T=0$.}
\end{center}
\end{figure}

In Fig.~\ref{f:beta}(a) we compare the HF position space distortion against the analytic expression (\ref{e:betagaussian}), and Hartree results. 
For all results, $\beta$ decreases with increasing dipole interaction strength, i.e.~the gas distorts so as to extend along the direction that the dipoles are polarized. 
This distortion increases the number of dipoles in the energetically favorable head-to-tail configuration. 
The numerical results terminate at finite DDI due to the system nearing mechanical instability (e.g.~see \cite{Bisset2011}). We observe that at any given value of $D_t$ the position distortion is greatest in near-spherical geometries. The analytic results are independent of whether particles are bosons or fermions, and predict a linear change in $\beta$ with the interaction strength.  The Hartree results are in  good agreement with the full HF solutions for the distortion. This demonstrates that the position space distortion arises from the direct interaction. 

The $T^{-5/2}$ scaling in the finite temperature analytic prediction \eqref{e:betagaussian} suggests that large magnetostriction effects will be obtained at low temperature. For bosons the critical temperature for condensation $T_c$ forms a lower bound for the applicability of our theory. 
For fermions our theory could be applied as $T\to0$, however the non-degenerate form of the distribution used to derive the analytic expression \eqref{e:betagaussian} will be unreliable below the Fermi temperature. In Fig.~\ref{f:beta}(b) we compare the finite temperature analytic results to the HF calculations for a range of temperatures below the Fermi temperature. These results indicate that the analytic approximation \eqref{e:betagaussian} significantly overestimates the distortion for $T\lesssim 0.5T_F^0$, and diverges as $T\to0$. We emphasize that this breakdown is not entirely due to the first order expansion used to derive the analytic results, but also because the distribution  \eqref{e:WT} fails to work in the degenerate regime. The $T=0$ analytic expression \eqref{e:betastep} is also shown in Fig.~\ref{f:beta}(b) and provides a qualitative estimate of the HF calculations at $T\to0$. However, for the interaction strength used in Fig.~\ref{f:beta}(b)  ($D_t=1$) corrections to our linearized result are apparent. In the inset to Fig.~\ref{f:beta}(b)  we compare the HF results to the analytic approximation at $T=0$ showing that good agreement is obtained for smaller values of $D_t$.

\begin{figure}[!tH]
\begin{center} 
\includegraphics[width=3.4in]{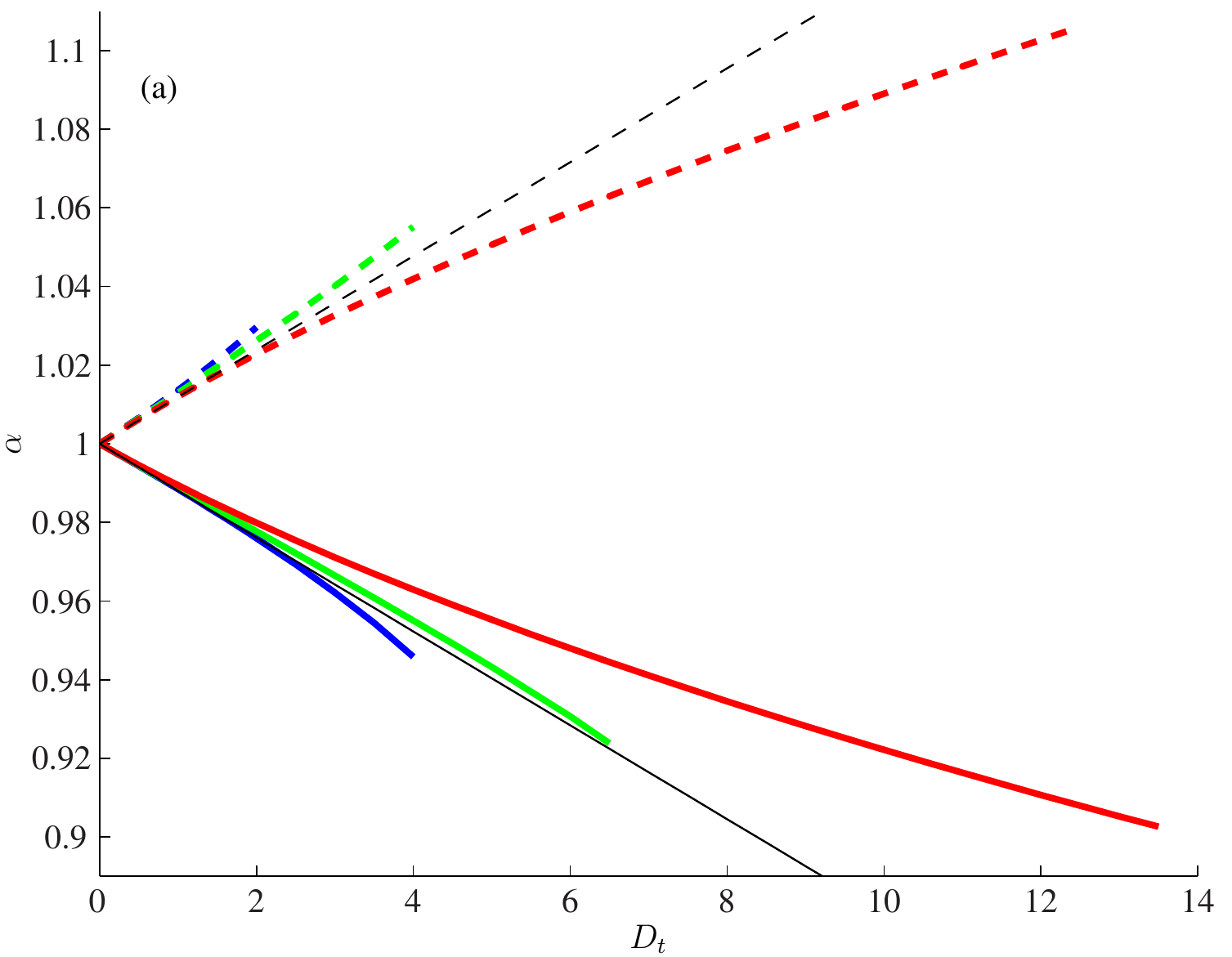} 
\includegraphics[width=3.4in]{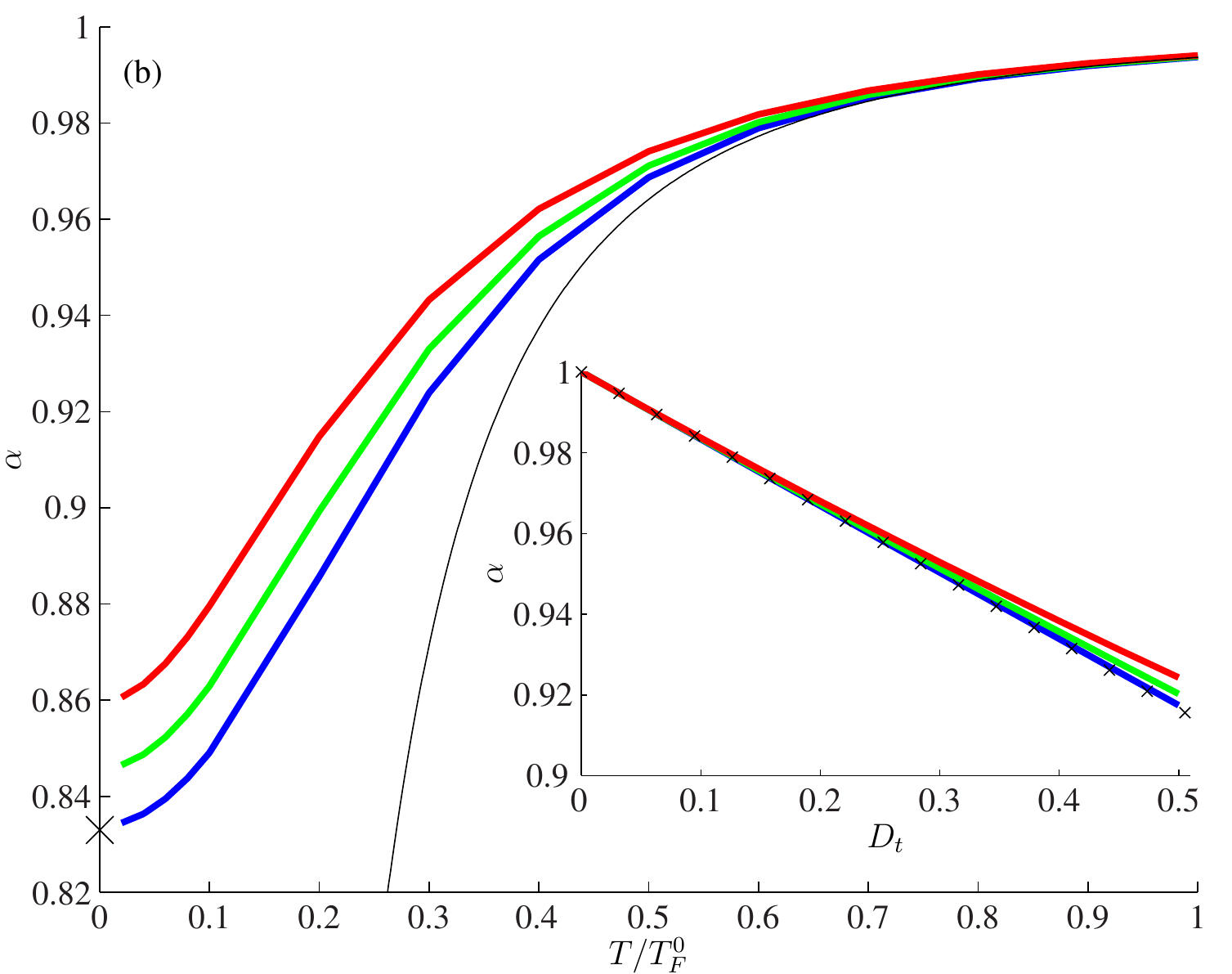}  
\caption{\label{f:alpha}(Color online) Momentum distortion for fermions (solid curves) and bosons (dashed curves) and the predictions from \eqref{e:alphastep} (black cross) and \eqref{e:alphagaussian} (thin black curves) with  $\lambda$ = 0.1 [blue (dark gray)], 1 [green (light gray)], and 10 [red (gray)]. (a) Bosons and fermions at $T=1.5T_c^0\approx0.78T_F^0$, (b) Fermions with $D_t=1$, (inset) Fermions at $T=0$.}
\end{center}
\end{figure}

\subsection{Momentum space magnetostriction}
In Fig.~\ref{f:alpha}(a) we compare the full HF results for the momentum space distortion against the analytic expression (\ref{e:alphagaussian}). For the fermionic system $\alpha$ decreases with increasing dipole interaction strength, i.e.~the gas distorts in momentum space to increase its extent in the  $k_z$ direction. In contrast, for the bosonic system, $\alpha$ increases with increasing dipole interaction strength, thus the system contracts along the $k_z$ direction relative to its radial extent. The Hartree theory (not shown) predicts no momentum distortion (i.e.~$\alpha=1$), thus momentum distortion arises solely from the exchange interactions, which are neglected in the Hartree approximation. As in the previous section, the numerical results terminate at finite DDI due to the system nearing mechanical instability.

In Fig.~\ref{f:alpha}(b)  we consider the Fermi system in the low temperature regime ($T\lesssim T_F^0$). Similar to the position space distortion [see Fig.~\ref{f:beta}(b)] we observe that the analytic approximation \eqref{e:alphagaussian} significantly overestimates the distortion for $T\lesssim 0.5T_F^0$, and diverges as $T\to0$. The $T=0$ analytic expression \eqref{e:alphastep} provides a qualitative estimate of the HF calculations at $T\to0$. In the inset we compare the HF results to the analytic approximation at $T=0$ showing that good agreement is obtained for smaller values of $D_t$.

\subsection{Exchange interactions and second order correlations}\label{SecDirectExchange}
To understand the role of exchange effects it is useful to consider the second order correlation function
\begin{align}
G^{(2)}(\mathbf{x},\mathbf{x}')=\left\langle \hat{\psi}^\dagger(\mathbf{x}) \hat{\psi}^\dagger(\mathbf{x}') \hat{\psi}(\mathbf{x}')\hat{\psi}(\mathbf{x})   \right\rangle,
\end{align}
which describes the probability of finding particles at positions $\bx$ and $\bx'$, with $\hat{\psi}(\bx)$ being the field operator. It is convenient to transform this to center of mass $\mathbf{R}=\tfrac{1}{2}(\bx+\bx')$ and relative $\mathbf{r}=\bx-\bx'$ coordinates, i.e.~$G^{(2)}(\mathbf{R},\mathbf{r})$, and introduce the  pair correlation function
\begin{equation}
C(\mathbf{R},\mathbf{r})= {G^{(2)}(\mathbf{R},\mathbf{r})} -n(\mathbf{R}+\tfrac{1}{2}\mathbf{r})n(\mathbf{R}-\tfrac{1}{2}\mathbf{r}).
 \end{equation}
Within HF theory, higher order correlation functions decompose into products of first order correlation functions (e.g.~see \cite{Naraschewski1999a,Baillie2010b}) giving
\begin{equation}
C(\mathbf{R},\mathbf{r}) = {\eta} \left|\int \frac{d\bk}{(2\pi)^3}W(\mathbf{R},\bk)e^{i\bk\cdot\mathbf{r}}\right|^2.\label{cHF}
\end{equation}
The $\eta$ prefactor of this result shows the important effect of quantum statistics on correlations: For fermions $C(\mathbf{R},\mathbf{r})\le0$ indicates their tendency to anti-bunch, while for bosons $C(\mathbf{R},\mathbf{r})\ge0$ indicates their tendency to  bunch \cite{Jeltes2007a}. Around any given point $\mathbf{R}$ the pair correlations decay as $|\mathbf{r}|$ increases, with this decay length defining the  correlation length.  Equation (\ref{cHF}) shows that the $\mathbf{r}$ dependence of the correlation function is related to the Fourier transform of the momentum distribution. Thus anisotropy in the momentum dependence of $W(\mathbf{R},\bk)$ will be accompanied by an inverse anisotropy in the  $\mathbf{r}$ dependence of the pair correlation function.

\begin{figure}[!t]
\begin{center} 
\includegraphics[width=3.4in]{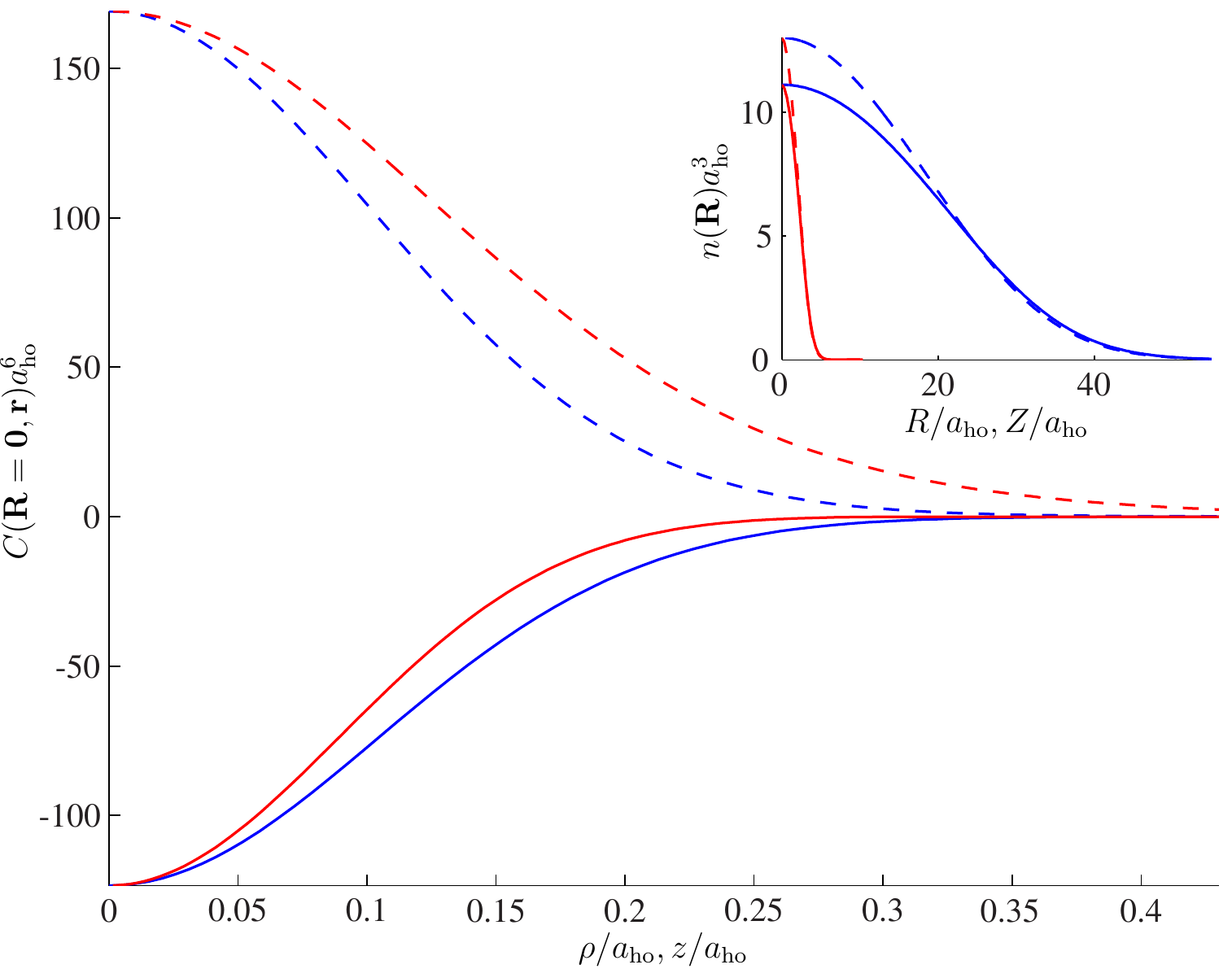} 
\caption{\label{f:g2} (Color online) Second order correlation function for $\lambda=10$, $D_t=4$, $N=10^5$, $T=1.1T_c^0\approx 0.57T_F^0$ for Fermi (solid curve) and Bose (dashed curve) for $\br=\rho\hat{\rho}$ (blue, dark grey) $\br=z\hat{z}$ (red, light grey). Inset shows the density profiles for comparison.  }
\end{center}
\end{figure}

We present a comparison of $C(\mathbf{R},\mathbf{r})$ for a Bose and Fermi gas in Fig.~\ref{f:g2} that demonstrates the features discussed above. The inset to that figure shows the density distribution, and reveals that the correlation length scale ($\sim\lambda_{\mathrm{dB}}$) is much smaller than the spatial extent of the systems.

In terms of the density and pair correlation functions we can evaluate the interaction energy densities
\begin{align}
\mathcal{E}_D(\mathbf{R})&=\frac{1}{2}\int d\mathbf{r}\,\UD(\mathbf{r})\,n(\mathbf{R}+\tfrac{1}{2}\mathbf{r})n(\mathbf{R}-\tfrac{1}{2}\mathbf{r}),\label{EDden}\\ 
\mathcal{E}_E(\mathbf{R})&=\frac{1}{2}\int d\mathbf{r}\,\UD(\mathbf{r})\,C(\mathbf{R},\mathbf{r}),\label{EEden}
\end{align}
for the direct and exchange terms, respectively. Note that the total direct and exchange energies, obtained by integrating the respective energy densities, are given by
\begin{align}
E_D&=\int d\mathbf{R}\,\mathcal{E}_D(\mathbf{R})=\frac{1}{2}\!\int \frac{d\bx d\bk}{(2\pi)^3} \Phi_D(\bx)W(\bx,\bk),\\
E_E&=\int d\mathbf{R}\,\mathcal{E}_E(\mathbf{R})=\frac{\eta}{2}\!\int \frac{d\bx d\bk}{(2\pi)^3} \Phi_E(\bx,\bk)W(\bx,\bk).\label{EE}
\end{align}
Equation (\ref{EDden}) shows that the direct energy arises from the density distribution of the system, while the exchange term (\ref{EEden}) arises from correlations, i.e.~with the integrand in Eq.~(\ref{EEden}) only being non-zero over the small range of $\mathbf{r}$ values for which correlations extend   [e.g.~see Fig.~\ref{f:g2}]. 

The anisotropy of the correlations ensures that the exchange energy is minimized. To see this we note  that the angular part of $\UD(\br)$ is proportional to the $Y_2^0(\theta,\phi)$ spherical harmonic, so that the integral in Eq.~(\ref{EEden}) is sensitive to a quadrupolar moment (i.e.~distortion) of $C(\mathbf{R},\mathbf{r})$ in the $\mathbf{r}$ variable.  For bosons, where the correlation is positive, the correlation peak is prolate:   short range correlations increase the number of pairs in the energetically favorable head-to-tail configuration. In fermions, where the correlation is negative, the correlation dip is oblate:    short range correlations reduce the number of pairs in the energetically unfavorable side-by-side configuration.

Finally, while our focus here has been on second order correlations, because of their relationship to energy, the exchange effects are also apparent in the first order coherence function $G^{(1)}(\bx,\bx')\equiv\langle \hat{\psi}^\dagger(\bx')\hat{\psi}(\bx)\rangle$ which, in center of mass and relative coordinates, relates to the 
distribution as
\begin{equation}
G^{(1)}(\mathbf{R},\mathbf{r}) = \int \frac{d\bk}{(2\pi)^3}W(\mathbf{R},\bk)e^{i\bk\cdot\mathbf{r}}.\label{G1}
\end{equation}
  This function also exhibits the anisotropy from exchange interactions with a differing sense of distortion for Bose and Fermi gases.

\section{Discussion and Conclusions}

In this paper we have examined  magnetostrictive effects in normal Bose and Fermi gases with dipole-dipole interactions. 
We have performed  Hartree-Fock calculations for these systems and compared the results against Hartree calculations and analytic approximations that we have developed. Magnetostrictive effects are only apparent for moderately large dipole strengths ($D_t$ of the order of unity), as being realized in current experiments with polar molecules. E.g.~$N=10^5$ KRb molecules (for which  both Bose \cite{Aikawa2010a} and Fermi \cite{Ni2008a} molecules have been realized) with $d=0.5\:\mathrm{D}$ in a  $\omega=2\pi\times100$ s$^{-1}$ trap corresponds to $D_t=4.7$ (higher interaction strengths are possible with more polar molecules, e.g.~NaK which has $d\approx2.72\:\mathrm{D}$ \cite{Wu2012a}). We note that Hartree-Fock theory should provide an accurate description of the system in the weak-coupling regime, e.g.~in the regime of dipolar atom  (magnetic dipole) experiments. For polar molecule experiments, because of the larger dipole strengths, corrections to HF theory may become important, however it is expected that the HF results should be  qualitatively accurate (e.g.~see discussion in Ref.~\cite{Kestner2010a}, also see \cite{Chan2010a,*Liu2011a,*Wang2008a}).  Here we have focused on the case of pure DDIs to allow a more direct comparison of Bose and Fermi systems.  However, we make a few general comments about the effect of contact interactions on the Bose system based on calculations we have performed. First, repulsive contact interactions allow the normal Bose gas to be stable out to higher dipole strengths (e.g.~see \cite{Bisset2011}), where the DDI induced distortions are larger. Second, for a given case (i.e. number of particles, DDI strength and temperature) the addition of repulsive contact interactions tends to lower the density and this in turn causes the DDI induced distortions to slightly decrease.

It should be possible to experimentally verify the effects we have studied in this paper. Spatial distributions have been precisely measured using absorption imaging in cold atom experiments   \cite{Hung2011a,Hung2011b}, and the extension of these techniques to the dipolar gas should allow the position space magnetostrictive behavior to be characterized. The direct measurement of the momentum space effect is possibly more challenging. Releasing the system from the trap requires a model of the expansion dynamics (e.g.~see   \cite{Sogo2009a,Lima2010b,Zhang2011a}) and does not simply reveal the momentum behavior.   Bragg spectroscopy could be used to probe the momentum distribution \textit{in situ} \cite{Stenger1999a,Stamper-Kurn1999a}.  As discussed in this paper  the momentum space magnetostrictive effects in dipolar gases are revealed by the development of anisotropic correlation functions. These could be revealed by a measurement of the  second order (density-density) correlation function, or by measurements of the first order coherence function [c.f.~Eq.~(\ref{G1})]. Such measurements have been performed with high accuracy on alkali Bose gases   \cite{Hung2011a,Hung2011b,Donner2007a}.

\section*{Acknowledgments}  This work was supported by the Marsden Fund of New Zealand (contract UOO0924).
\appendix
\section{Analytic approximations}

\begin{figure}[!t]
\begin{center} 
\includegraphics[width=3.4in]{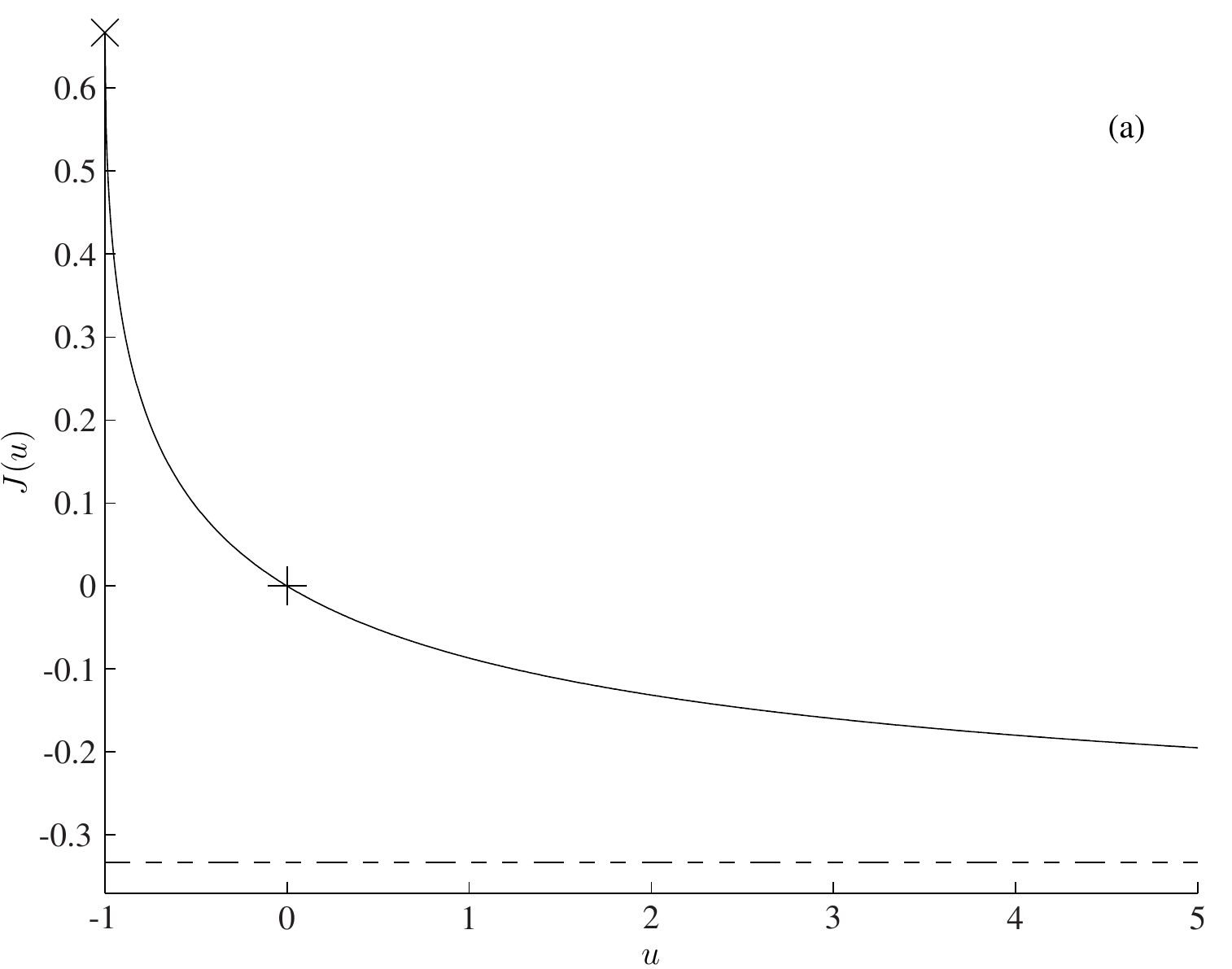} 
\includegraphics[width=3.4in]{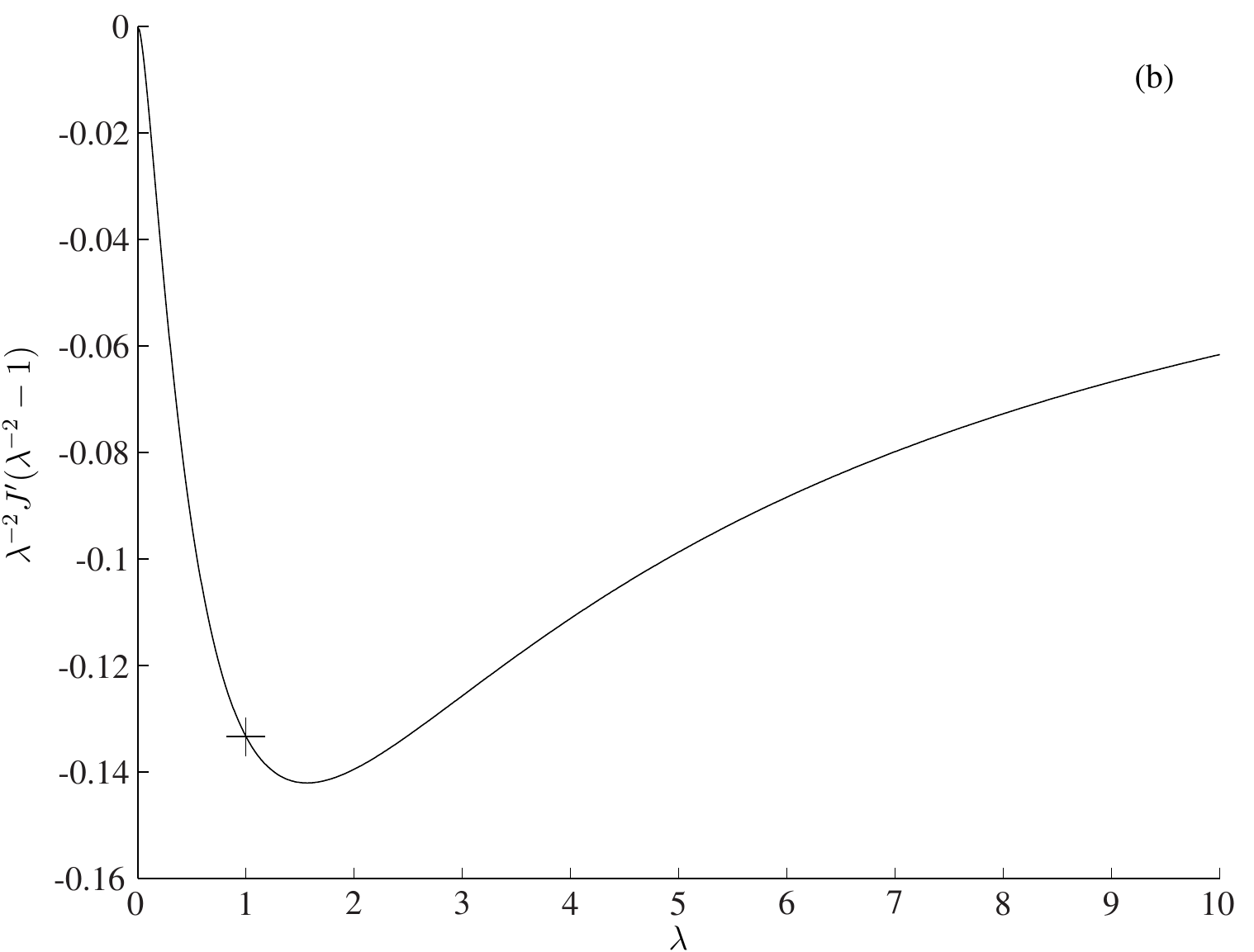}
\caption{\label{f:J} (a) The function $J(u)$ (solid curve) showing $J(-1)=\frac{2}{3}$ (cross), $J(0)=0$ (plus) $J(u)\to -\frac{1}{3}$ as $u\to\infty$ (dashed line). 
(b)  The term appearing in \eqref{e:betastep} and \eqref{e:betagaussian} as a function of the aspect ratio showing $J'(0)=-\frac{2}{15}$ at $\lambda=1$ (plus).}
\end{center}
\end{figure}

\subsection{Zero temperature fermions}
Using \eqref{e:W0}, the energy of a dipolar gas described by $W_0$ is (after \cite{Miyakawa2008a}), ignoring factors of $\chi_\rho^2\chi_z$ and $\kappa_\rho^2\kappa_z$ which do not contribute to the distortions to first order
\begin{align}
    \frac{\langle H\rangle_0}{\hbar\omega N^{4/3} } &= 
    \frac{3^{1/3}}{2^{5/3}}\left[\frac{1}{\kappa_\rho} + \frac{1}{2\kappa_z}+ \frac{1}{\chi_\rho} + \frac{1}{2\chi_z}\right] \notag\\
    &+\frac{2^{11}D_t}{3^{5/2}35 \pi^2}  \left[J\left(\frac{\chi_\rho}{\lambda^{2}\chi_z}-1\right) - J(\delta)\right], \label{e:ET0}
\end{align}

where $\delta = \kappa_z/\kappa_\rho-1$ and\footnote{We note that $(\sinh^{-1}\sqrt{u})/\sqrt{u}$ is real for $u\ge-1$ and that our $J(u)=I[(1+u)^{1/3}]/6$ of \cite{Sogo2009a}}
\begin{align}
    J(u) &\equiv \frac{1}{2} \int_0^{\pi} d\theta \sin\theta  \left(\frac{\cos^2\theta}{1 + u\sin^2\theta} - \frac13\right)\\
    &=    \frac{1}{u}\left[\sqrt{1+u} \frac{\sinh^{-1} \sqrt{u}}{\sqrt{u}} -1 \right] -\frac13.\label{e:Ju}
\end{align}
For very small values of $u$, the series expansion
\begin{align}
    J(u)  &=
    -\frac{2 u}{15}+\frac{8 u^2}{105}-\frac{16 u^3}{315}+ \od{u^4}
\end{align}
and its derivative are useful. We plot $J(u)$ in Fig.~\ref{f:J}(a) and note that it is monotonically decreasing and that it is zero in the spherical case, i.e. $J(0)=0$. 

To find $\alpha=\sqrt{1+\delta}$ and $\beta=\sqrt{\chi_z/\chi_\rho}$ to first order, we minimize the energy \eqref{e:ET0}, assuming $\delta$ and $\chi_z/\chi_\rho-1$ are small parameters, we find Eqs.~\eqref{e:alphastep} and \eqref{e:betastep}. We plot the direct interaction derivative term, $\lambda^{-2}  J'(\lambda^{-2}-1) $,  appearing in Eq.~\eqref{e:betastep} [also in Eq.~\eqref{e:betagaussian}], in Fig.~\ref{f:J}(b) and note that the rate of change of direct interaction energy is largest for spherical to slightly oblate geometries, so that these are most able to offset the increase in trap energy and are likely to distort most.

The derivative of $J(u)$ is required to evaluate Eqs.~\eqref{e:betastep} and \eqref{e:betagaussian} and is given explicitly by
\begin{align}
    J'(u) &= \frac{3  -(2 u+3) \frac{\sinh^{-1}\sqrt{u}}{\sqrt{u(u+1)}}}{2 u^{2} }.
\end{align}

\subsection{Finite temperature}
Using \eqref{e:WT}, the energy of a dipolar gas described by $W_T$ is (after \cite{Sogo2009a,Endo2011a}), ignoring factors of $\chi_\rho^2\chi_z$ and $\kappa_\rho^2\kappa_z$ as above  
\begin{align}
    \frac{\langle H\rangle_T}{N\kt} &= \frac{1}{\kappa_\rho} + \frac{1}{2\kappa_z}+ \frac{1}{\chi_\rho} + \frac{1}{2\chi_z}\notag\\
    &+  \frac{D_t}{4\sqrt{\pi}} \left(\frac{\hbar\omega N^{1/3}}{\kt}\right)^{5/2} \left[J\left(\frac{\chi_\rho}{\lambda^{2}\chi_z}-1\right) +\eta J(\delta)\right]. \label{e:ET}
   \end{align}

Minimizing the energy \eqref{e:ET} to first order (other terms in the free energy do not depend on $\delta$ or $\chi_z/\chi_\rho$ directly, so do not contribute at this level of approximation) we find Eqs.~\eqref{e:alphagaussian} and \eqref{e:betagaussian}.

%

\end{document}